# Hierarchically porous Ni monolith@branch-structured $NiCo_2O_4$ for high energy density supercapacitors

Qin Guo*, Mengjie Xu, Rongjun Xu, Ying Zhao, Boyun Huang*

*State Key Laboratory of Powder Metallurgy, Central South University, Changsha, Hunan, P. R. China 410083.*

Email: hby@mail.csu.edu.cn. guoqin999@gmail.com

**Abstract**: $NiCo_2O_4$ of varying nanostrucutures ranging from nanowires, nanoplates to nanoplates@nanowires were successfully grown on microporous (MP) Ni foams via one-step hydrothermal process. The investigation of electrochemical capacitance favors $NiCo_2O_4$ of nanoplates@nanowires microstructures which possesses specific capacitance of 1380.3 F/g and 1033F/g at 5A/g and 50A/g respectively and 86.7% capacitance retention after 5000 cycles at 30A/g. The relationship between morphology and specific capacitance was further explored by the model of surface roughness factor (RF), which is indicative of the active electrode-electrolyte interface areas. The RF of porous Ni@$NiCo_2O_4$ was remarkably improved by employing hierarchically porous (HP) Ni monoliths as substrates, which illustrates the model of high energy density (12.6 F/$cm^2$) electrodes for super-capacitors.

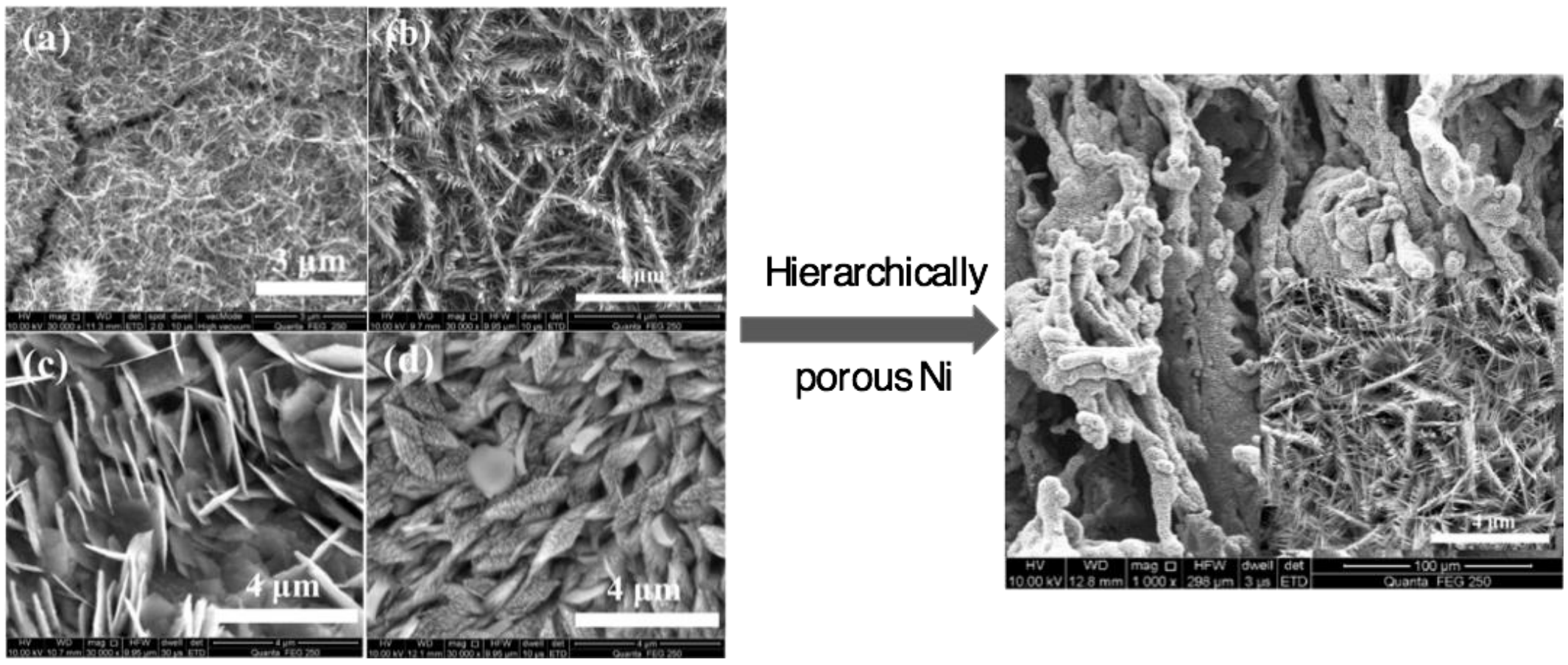

## Introduction

Electrochemical supercapacitors, offering extremely high power density and reasonably high energy density, are considered as one of the most important next generation energy storage devices.[1-3]As to the electrode materials, transition metal oxides (TMOs) have drawn extensive attention in recent years, owing to their multiple oxidation states that are desirable for pseudocapacitance generation. Among them, $RuO_2$ is a prominent candidate because of its superior conductivity and high specific capacitance of 1580 F $g^{-1}$.[4, 5]The commercialization of $RuO_2$ based supercapacitors, however, is not promising due to the high cost and rareness of Ru. Recently, more environment-friendly TMOs, such as NiO, $Co_3O_4$, $MnO_2$, $NiCo_2O_4$, $CoMn_2O_4$, have been intensively evaluated to achieve high-energy storage capacity at lower costs.[6-11] Specifically, spinel nickel cobaltite ($NiCo_2O_4$), which possesses a much better electronic conductivity and superior electrochemical activity than those of nickel oxides and cobalt oxides,[12] is a promising cost-effective alternative for expensive $RuO_2$.

Since pseudo-capacitance is based on faradic redox reactions which occur on the very surface and sub-surface regions of electrodes, high specific surface area and optimized pore structure of the electrode materials are critical to both exert the potential of materials and facilitate the mass transfer of electrolytes.[13-15] There has been two trends to achieve the double-win goal. On one hand, $NiCo_2O_4$ with various nanostructures such as flower-like,[16, 17] urchin-like,[18, 19] porous network-like,[20] nanowires,[21, 22] and nanoflakes,[20, 23, 24] by well-developed morphology controlled synthesis have been investigated as energy storage devices. One the other hand, 2D metal foils has been replaced by 3D porous conductive frames (such as Ni foams, carbon cloth) as substrates to directly grow $NiCo_2O_4$ materials and construct high perfor-mance super-capacitor electrodes with improved areal specific capacitance.[25-30] Although comprehensive enhancement of electrochemical performance has been achieved in $NiCo_2O_4$ nano-structured electrodes, the relationship between microstructures and electrochemical properties has not been well understood. Few breakthroughs have been made regarding the optimized conductive substrates either.

In this work, $NiCo_2O_4$ with various microstructures ranging from nanowires, nanoplates and branch-structures were successfully grown on microporous Ni foams via one-step hydrothermal process. Electrochemical performance of different nano-structures were comprehensively investigated. Electrochemical double-layer capacitance (EDLC) based surface roughness factor (RF) model was applied to understand the relationship between microstructures and electrochemical capacitance. Furthermore, hierarchically porous Ni was employed as replacement of microporous Ni foams and noticeably high specific areal capacitance and energy density was achieved. It is anticipated that this work should shed light on the design and development of next generation high performance electrochemical supercapacitors.

**Experimental section**

*Materials synthesis.* $NiCo_2O_4$ with different microstructures were grown on porous Ni substrates by hydrothermal (HT) method followed by a heat treatment. The reagents of analytical grades were used as purchased. In a typical case, 2 mmol $Ni(NO_3)_2{\cdot}6H_2O$, 4 mmol $Co(NO_3)_2{\cdot}6H_2O$ , and 24 mmol urea were dissolved in 80 ml deionized water. After the addition of 6 mmol $NH_4F$, the solution was transferred into a 100 mL Teflon-lined stainless steel autoclave. Then discs of microporous (MP) Ni foams (420g cm$^{-2}$, Changsha Liyuan New Material Co., Ltd.) with diameter of 16 mm and mass of ~85mg were added into the solution as growth substrates. Afterwards, the autoclaves were sealed and heated in an oven at 100 °C for 5h, and then cooled naturally to room temperature. After being washed sufficiently and ultra-sonicated for 2 min, the MP Ni foams with purple precursors were annealed in air flow at a heating rate of 2 °C/min to 300 °C for 2 h. The mass increase compared with the porous Ni substrates was taken as the mass of electrochemically active $NiCo_2O_4$. $NiCo_2O_4$ with different microstructures but similar mass were obtained by varying combination of $NH_4F$ dosages and hydrothermal reaction time with (2 mmol, 10h) for nanowires (NWs), (6 mmol, 5h) for vertical-nanoplates@nanowire (V-NPs@NWs), (12 mmol, 3h) for vertical-nanoplates (V-NPs), and (24 mmol, 1h) for horizontal-nanoplates (H-NPs). To evaluate the influence of substrate microstructures, MP Ni foams were replaced by hierarchically porous (HP) Ni frames of equivalent size and mass synthesized by combustion method.[31]

*Materials characterization.* The pyrolysis property of precursors was analyzed by thermal gravimetric−differential scanning calorimetry (TG–DSC, Netzsch STA449C Jupiter) in air at a heating rate of 10℃/min. The structural information of precursors and annealed products was characterized with X-ray powder diffractometer (Rigaku D/max-2550, Cu Kαradiation, λ= 1.5406 Å). The chemistry of the annealed products was tested by Inductively Coupled Plasma-Atomic Emission Spectrometry (ICP-AES, Thermo Jarrell Ash IRIS Advantage 1000). Scanning electronic microscopy (SEM) micrographs were acquired by FEI Qanta FEG 250 microscope operated at 10 kV. The crystal structure and microstructure have been further characterized by TEM (JOEL 2100F). The specific surface area was characterized by low temperature $N_2$ adsorption and desorption method (Quadrasorb SI-3MP).

*Electrochemical performance.* The as prepared porous Ni@$NiCo_2O_4$ were tested as supercapacitor electrode in 3M KOH by a three electrodes cell connected to Arbin battery testing system(S/N 173123-A) with Pt as counter electrode and Hg/HgO as referring electrode. Firstly, surface roughness factor (RF) of the electrodes was characterized by cyclic voltammetry (CV) in voltage window of 0.05 to 0.1 vs Hg/HgO where only electrochemical double-layer capacitance was generated. Then, the electrodes were electrochemically activated with CV between 0 to 0.6V v.s. Hg/HgO at a scanning rate of 20 mV/s for 1000 cycles. Afterwards, CV was recorded between 0 to 0.6V v.s. Hg/HgO at different scanning rates. Galvanostatic charge−discharge (GCD) tests were conducted between 0 to 0.55 V v.s. Hg/HgO to investigate the rate performance and then cycling stability. The gravimetric specific capacitance $C_m$, areal specific capacitance $C_a$, energy density E and power density P were calculated according to the following equations

$$C_m = \frac{I\Delta t}{m\Delta V} \text{ and } C_a = \frac{I\Delta t}{A\Delta V}$$

$$E = \frac{1}{2} \times C \times \Delta V^2$$

$$P = \frac{E}{\Delta t}$$

Where *m* is the mass of active materials, *A* is the apparent area of electrodes, *I* is the discharging current, $\Delta t$ is the discharging time and $\Delta V$ is the voltage drop during discharging.

## Results and discussion

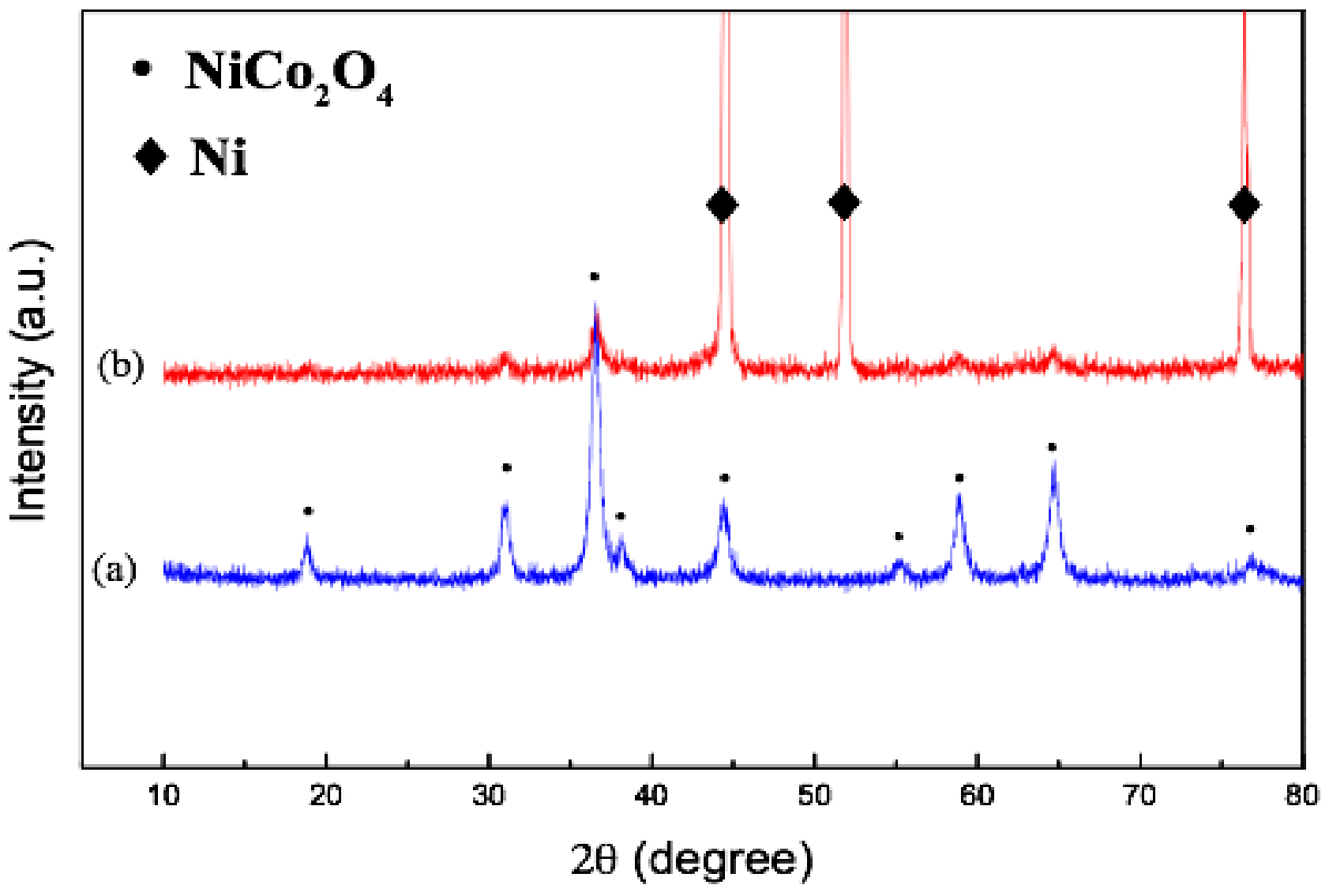

Fig. 1 XRD patterns of (a) $NiCo_2O_4$ powders (b) MP Ni@$NiCo_2O_4$

The synthesis of $NiCo_2O_4$ consists of the precipitation and phase transformation of precursors. Firstly, carbonate hydroxides of (Ni, Co) were obtained according to reactions shown in the Experimental Section of Electronic Supporting Information (ESI). As shown in the XRD patterns of precursors (Fig. S1), the broad and weak diffraction peaks can be assigned to $Ni_2(OH)_2CO_3{\cdot}4H_2O$ (JCPDS no. 38-0714) and $Co(CO_3)_{0.5}(OH){\cdot}0.11H_2O$ (JCPDS no.48-0083), respectively.[32] The pyrolysis property of precursors was characterized by TG-DSC (Fig. S2). Decomposition of the as-prepared precursors was observed at around 300 ℃ and then the mass ratio remains constant after 400 ℃. To facilitate the decomposition of precursors but avoid the oxidization of porous Ni substrates, heat treatment was performed at 300℃ for 2h. After pyrolysis and oxidization of precursors (seen the Experimental Section of ESI), $NiCo_2O_4$ was obtained as shown in the XRD patterns in Fig. 1a. The diffraction peaks at around 19°，31°，36°，45°，59°，65° and 78° can be assigned to (111)，(220)， (311)，(400)，(511)，(440) and (531) planes of $NiCo_2O_4$ phase (JCPDS No.20-0781) respectively. However, the peaks of $NiCo_2O_4$ on Ni substrate is extremely weak due to relatively low mass ratio compared to Ni phase (see Fig. 1b). Considering the similar crystal structure of $Co_3O_4$ and $NiCo_2O_4$, further chemical analysis by ICP-AES was performed. The atom ratio of Co to Ni is ~2.4, suggesting the synthesis of $NiCo_2O_4$.

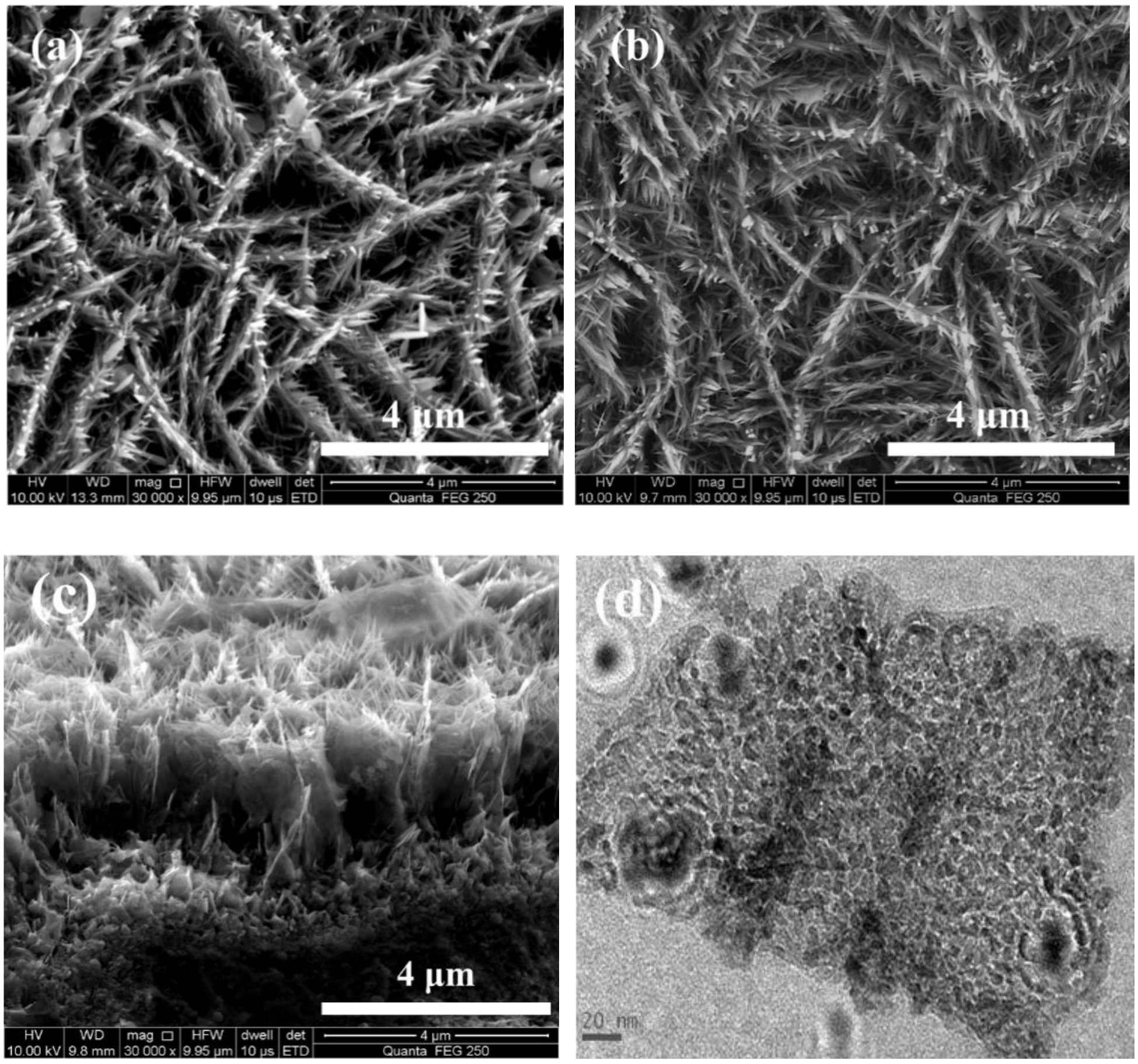

Fig. 2 SEM micrographs of MP Ni@ $NiCo_2O_4$ (a) before annealing and (b) (c) after annealing and (d) TEM micrographs of $NiCo_2O_4$ after annealing.

The morphology of precursors and $NiCo_2O_4$ products was characterized by SEM and TEM. As shown in Fig. 2b-c, branch-structured $NiCo_2O_4$ arrays have been synthesized by one-step hydrothermal method. The hierarchical structure composed of interconnected nanoplate arrays and nanowire branches extending into the space between. The morphology remains unchanged compared with the precursors (Fig. 2a). However, numerous mesopores were developed after heat treatment, as shown by the TEM micrographs (Fig. S 3 and Fig. 2d). Moreover, the specific surface area of $NiCo_2O_4$ increased significantly to 66.6 $m^2/g$ compared with that of precursors (21.4 $m^2/g$), which is favorable for increasing the interfaces between $NiCo_2O_4$ and electrolytes.

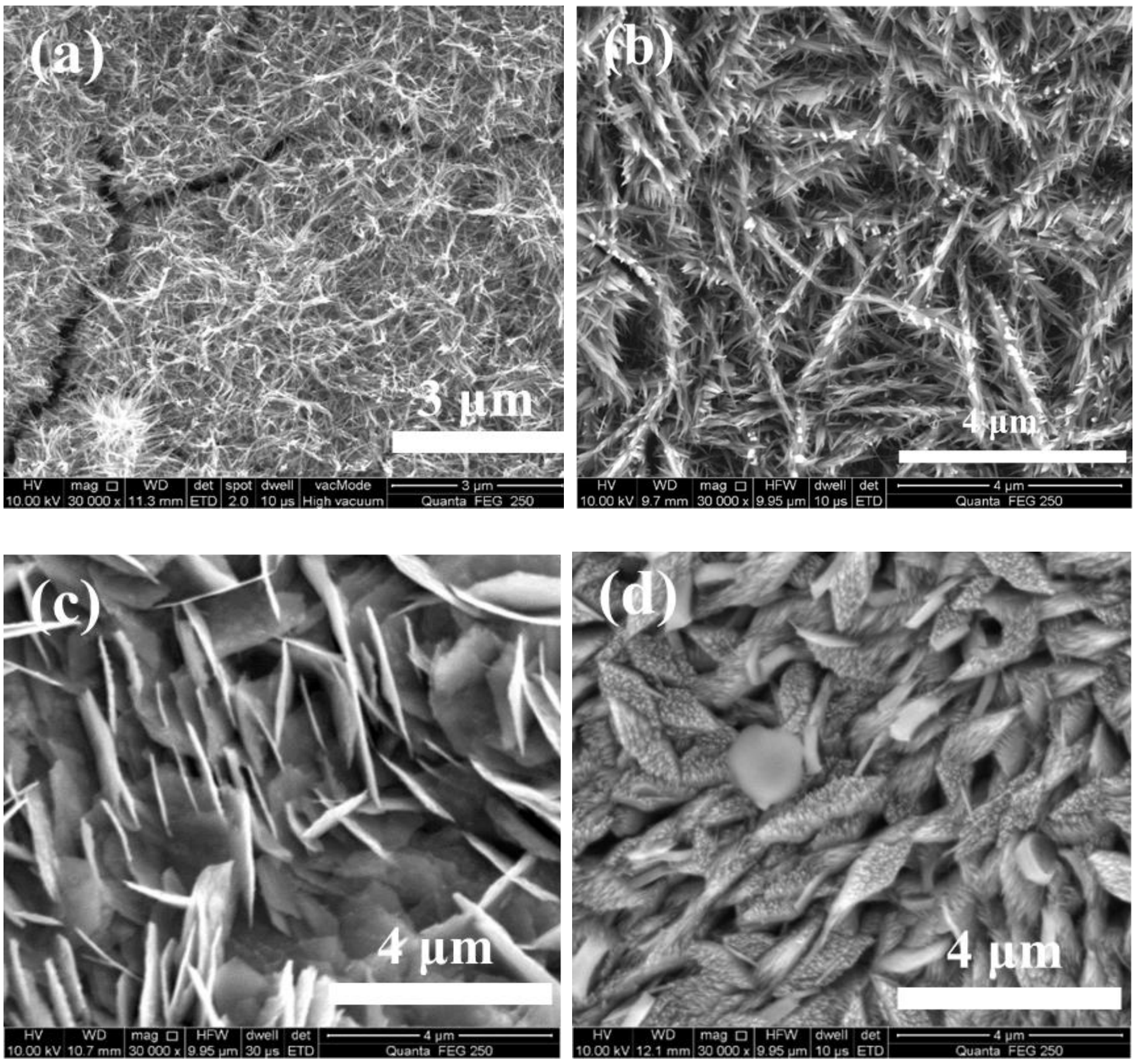

Fig. 3 SEM micrographs of $NiCo_2O_4$ synthesized with varying combination of $NH_4F$ dosages and hydrothermal reaction time (a) (2 mmol, 10h) (b) (6 mmol, 5h) (c) (12 mmol, 3h) (d) (24 mmol, 1h).

$NiCo_2O_4$ materials with various morphologies were obtained by adjusting the dosages of $NH_4F$ and reaction time. With the increase of $NH_4F$, the nanostructures evolve from nanowires (NWs) to vertical nanoplates@nanowires (V-NPs@NWs) and then horizontal nameplates (H-NPs)(see Fig. 3a, b and d, respectively). Vertical nanoplate(V-NPs) arrays have been synthesized by lowering the reaction time to 3h (see Fig. 3c). $NH_4F$ acts as an activating agent to the Ni substrates, promoting the formation of nanoplates structures.[33] The interplay of the dosages of $NH_4F$ and reaction time results in a variety of morphologies observed in $NiCo_2O_4$ materials.

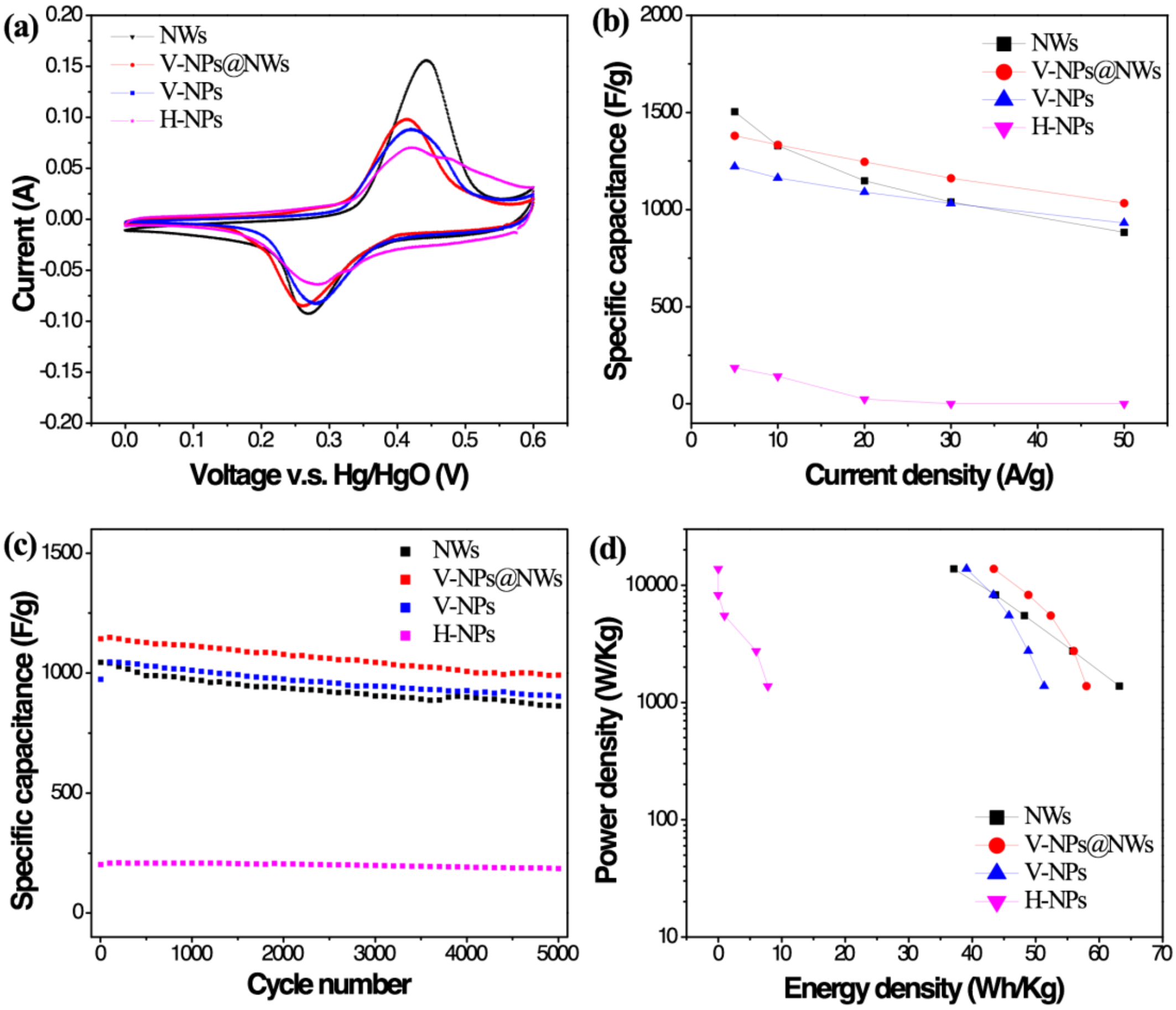

Fig. 4 Electrochemical performance of $NiCo_2O_4$ with different microsturctures (a) CV at a scanning rate of 5 mV/s; (b) rate performance at current densities of 5 , 10 , 20 , 30 and 50A/g with a voltage window of 0~0.55V vs. Hg/HgO; (c) cyclic capacitance at current density of 30A/g with a voltage window of 0~0.55V vs. Hg/HgO; (d) Ragone plot.

The electrochemical performances of different $NiCo_2O_4$ nanostructures were compared and shown in Fig. 4. As can be seen from the CV test (Fig. 4a), symmetric anodic and cathodic peaks suggest the reversible redox reactions between MO/MOOH (M=Ni, Co). The GCD rate performance (Fig. 4b) for all microstructures shows a downward trend due to electron and ion transfer barrier at high current density.[34] Specifically, the $C_m$ of H-NPs remains constantly lower than other microstructures while the $C_m$ of NWs exhibits a sharp decrease from the highest 1503.7 F/g to 882.9 F/g with the increase of current density. Although the $C_m$ of V-NPs@NWs is the second highest as 1380.3 F/g at 5A/g, it remains 1033F/g, leading to the highest at 50A/g and granting it the best overall power density-energy density property (see Figure 4d). The V-NPs@NWs also exhibits excellent cycling stability with $C_m$ of 1143.5 F/g at 30 A/g, remaining 86.7% of the initial values after 5000 cycles (Fig. 4c). The morphologies of the cycled electrodes were evaluated using SEM, as shown in Fig. S 4. Cracking and dissolution of active materials may account for the degradation of $C_m$ for all microstructures above.

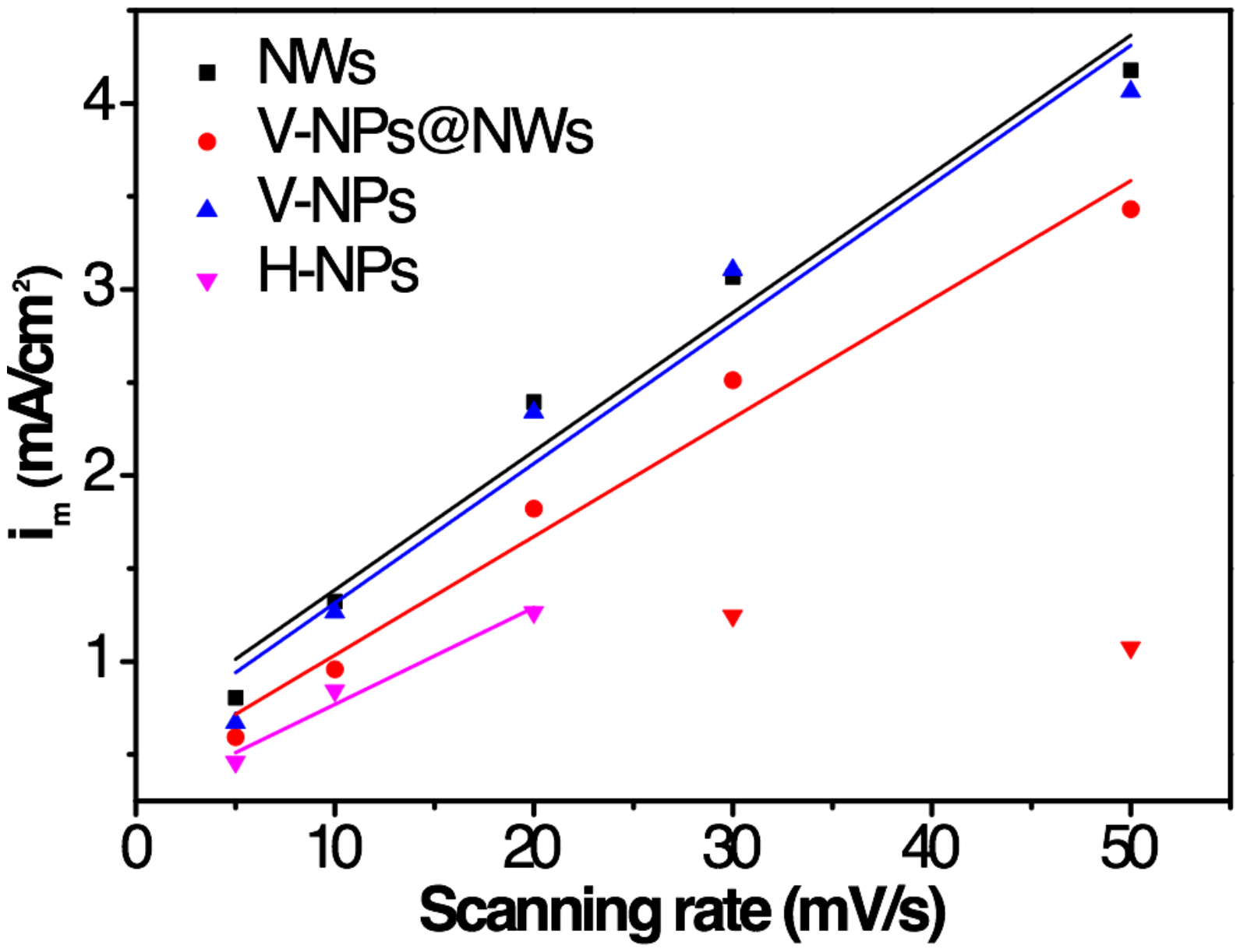

Fig. 5 The $i_m$-scanning rate relation for $NiCo_2O_4$ with different microstructures tested by CV at different scanning rates

To further understand the morphology-capacitance relationship, surface roughness factor (RF) of different $NiCo_2O_4$ nanostructures was estimated on the basis of double-layer capacitance using CV test (Fig. S 5). Double-layer capacitance ($C_{dl}$) of electrodes was calculated from the slope of $i_m$-scanning rate relations (Fig. 5), where $i_m$ is the mid-point (0.075V v.s. Hg/HgO) current normalized by the apparent electrode area and $\partial V/\partial t$ is scanning rate. RF was obtained by $C_{dl}$ divided by $C_0$ , where $C_0$ is the areal specific capacitance for smooth surface (60 μF/$cm^2$).[35] The calculated RFs for NWs, V-NPs@NWs, V-NPs and H-NPs are 1242.5, 1062.5, 1249.0 and 867. 2 respectively. It is interesting to note higher RFs for NWs and V-NPs than V-NPs@NWs even though V-NPs@NWs showed the best overall capacitance. In the case of capacitance test by CV, RF values are indicative of the active area of electrode-electrolyte interfaces with wetting condition considered. It is noteworthy that hierarchical branched-structures tend to make the surface more hydrophobic by pinning the droplet or forming air pockets, which leads to reduced solid–liquid interface area/smaller RF.[36] After long activation process of electrodes, the interface areas could be recovered by gradually removing wetting hysteresis and then promote better pseudo-capacitance performance.[36]

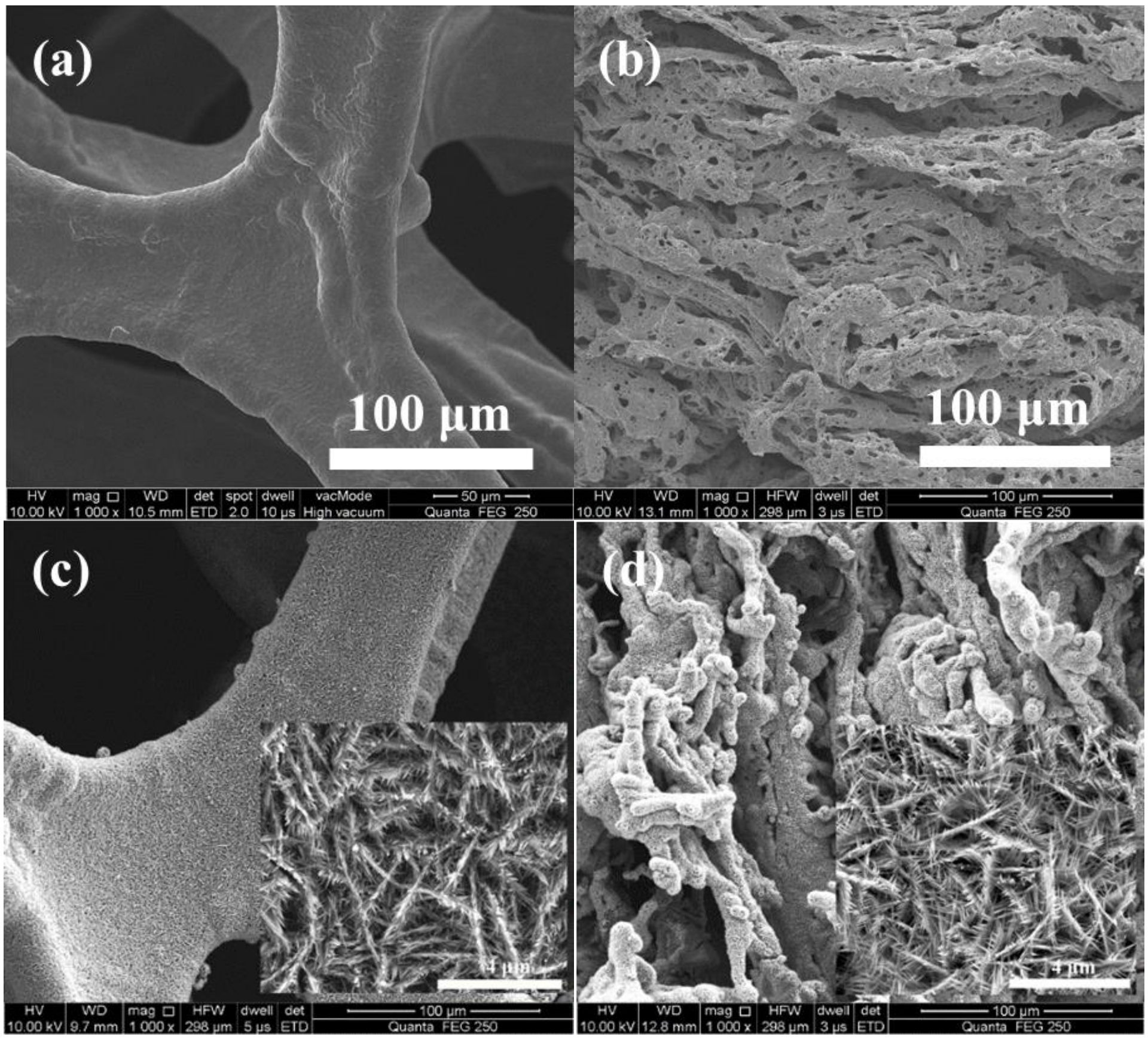

Fig. 6 SEM micrographs of MP Ni foam and HP Ni monoliths before (a) (b) and after (c) (d) $NiCo_2O_4$ growth.

To obtain larger interface areas, MP Ni foams were substituted by HP Ni monoliths[31] as the conductive substrates for $NiCo_2O_4$ growth (see Fig. 6a-b). The XRD patterns of both products can be assigned to Ni@$NiCo_2O_4$ (Fig. S 6). As depicted on Fig. 6c-d, V-NPs@NWs $NiCo_2O_4$ arrays grew uniformly on both substrates, demonstrating its transferability. However, much higher mass loading of $NiCo_2O_4$ (17.8 ~ 20.6 mg) was achieved for HP Ni frames than that for with MP Ni foams (3.2 ~ 4.4 mg) under the same synthesis condition. This can be ascribed to the optimized pore structure (2nm to 400μm) and higher specific surface area (5.63 $m^2/g$) of HP Ni monoliths compared with macro-porous structure (450~3200μm) and low specific surface area (~0.8 $m^2/g$) of MP Ni foams.

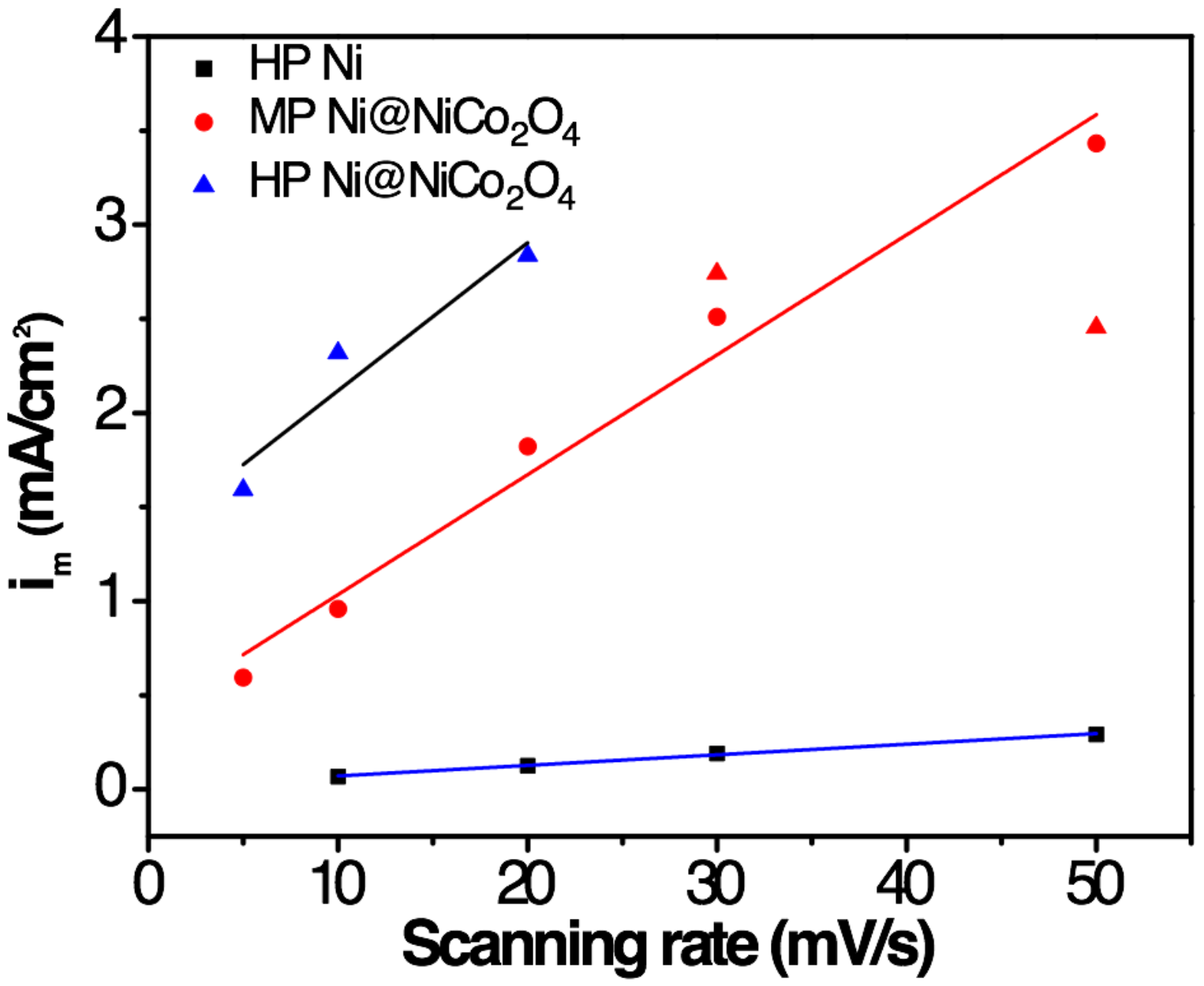

Fig. 7 The $i_m$-scanning rate relation for MP Ni, MP Ni@NiCo2O4 and HP Ni@$NiCo_2O_4$

The RF of $NiCo_2O_4$ arrays on different Ni substrates has been calculated based on $i_m$-scanning rate relation (Fig. 7). The calculation of RF for MP Ni failed because of the low current response intervened by background noise (see Fig. S 7). On contrast, the nake HP Ni gives RF of 93.8, which promotes high RF of 1309.9 for HP Ni@$NiCo_2O_4$.

Capacitance dispersion was observed for HP Ni@$NiCo_2O_4$ and H-NPs (Fig. 5) due to geometry aspect of rough electrodes or atomic scale inhomogeneities.[37] Althogh singular points in $i_m$-scanning rate relation were masked, the RF for HP Ni@$NiCo_2O_4$ is still believed to be underestimated due to wetting hysteresis.

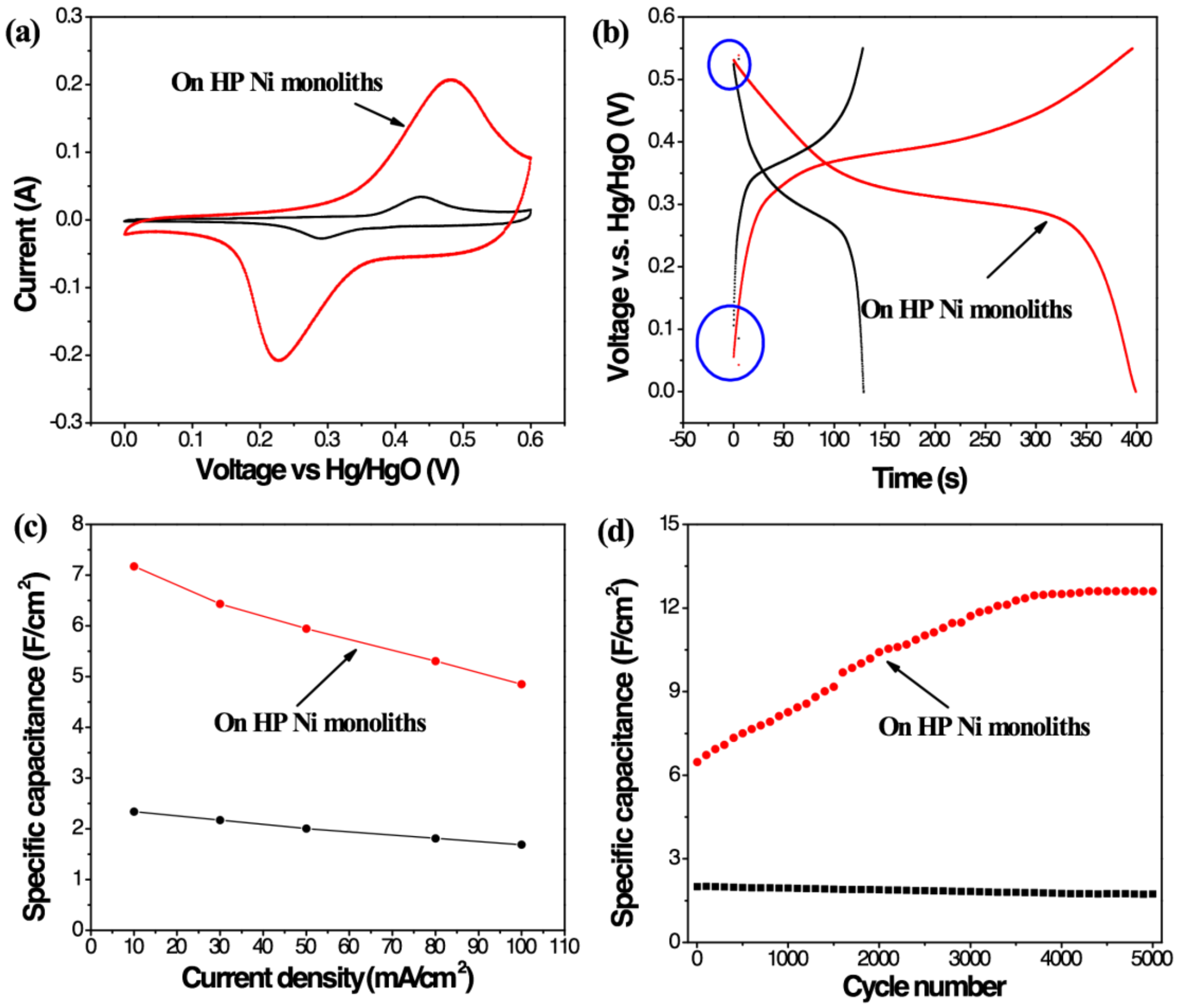

Fig. 8 Electrochemical performance of $NiCo_2O_4$ on different substrates (a) CV at a scanning rate of 5 mV/s; (b) GCD curves at a current density of 10 mA/cm$^2$; (c) rate performance at current densities of 10，30，50，80 and 100 mA/cm$^2$ between 0 and 0.55V v.s. Hg/HgO; (d) cyclic capacitance at a current density of 50 mA/cm$^2$ between 0 and 0.55V v.s. Hg/HgO (black line for MP Ni@ $NiCo_2O_4$)

The electrochemical performances of branch-structured $NiCo_2O_4$ on both porous Ni substrates were compared. As shown in **Fig. 8**a, CV curves with symmetrical peaks indicate the good reversibility of both electrodes. Howecer the response current for HP Ni@$NiCo_2O_4$ is significantly higher than that of MP Ni@$NiCo_2O_4$, indicating much larger capacitance, which can also be confirmed by the much longer discharging time (Fig. 8b). Also, lower IR drop/increase was observed after current inverse for HP Ni@$NiCo_2O_4$ due to much lower real current density (blue circle in Fig. 8b). Rate performance was further investigated with HP Ni@ $NiCo_2O_4$ and MP Ni@ $NiCo_2O_4$ exhibiting $C_a$ of 7.2 and 2.3 F/cm$^2$ at 10 mA/cm$^2$ respectively and 4.8 and 1.7 F/cm$^2$ at 100 mA/cm$^2$ respectively (Fig. 8c). Even much higher, only 66.7% capacitance retention was achieved for HP Ni@ $NiCo_2O_4$, indicating poor rate performance. The cycling stability of both electrodes was shown in Fig. 8d. The $C_a$ of HP Ni@$NiCo_2O_4$ increased from 6.5 F/cm$^2$ to 12.6 F/cm$^2$ while the $C_a$ of MP Ni@ $NiCo_2O_4$ dropped from 2.0 F/cm$^2$ to1.7 F/cm$^2$ after 5000 cycles at 50 mA/cm$^2$. The five-fold higher $C_a$ of HP Ni@$NiCo_2O_4$ may due to a long activation process when severe wetting hysteresis of the doubled hierarchical structures[36] was removed and a new record was established[25, 38-40] The $C_m$ of HP Ni@ $NiCo_2O_4$ based on overall mass reached 245.1 F/g at 50 mA/cm$^2$, which is even comparable with low density carbon substrates. [39] The branch-structured morphology for $NiCo_2O_4$ on HP Ni monolith remains unchanged after 5000 cycles while much dissolution was observed for $NiCo_2O_4$ on MP Ni foams(see Fig. S 8). The rate performance and cycling consistency may be improved by optimizing the wetting in solid-liquid interfaces.

**Conclusion**

In summary, $NiCo_2O_4$ with various nanostrucutures ranging from nanowires, nanoplates to nanoplates@nanowires were successfully grown on microporous (MP) Ni foams via one-step hydrothermal process. The investigation of electrochemical capacitance favors $NiCo_2O_4$ of nanoplates@nanowires microstructures which possesses specific capacitance of 1380.3 F/g and 1033F/g at 5A/g and 50A/g respectively and 86.7% capacitance retention after 5000 cycles at 30A/g. The relationship between morphology and specific capacitance was further explored by the model of surface roughness factor (RF), which is indicative of the active electrode-electrolyte interface areas. The RF of porous Ni@$NiCo_2O_4$ was remarkably improved by employing hierarchically porous (HP) Ni monoliths as substrates, which illustrates the model of high energy density (12.6 F/$cm^2$) electrodes for super-capacitors.

**Acknowledgements**

This work was supported by the Recruitment Program of Global Youth Experts, the National Natural Science Foundation of China (51304248), the Program for New Century Excellent Talents in University (NCET-11-0525), the Doctoral Fund of Ministry of Education of China (20130162110002), the Program for Shenghua Overseas Talents from Central South University and the State Key Laboratory of Powder Metallurgy at Central South University.

**Notes and references**

†Electronic supplementary information (ESI) available.

# Electronic Supplementary Information

---

## Hierarchically porous Ni monolith@branch-structured $NiCo_2O_4$ for high energy density supercapacitors

Qin Guo*, Mengjie Xu, Rongjun Xu, Ying Zhao, Boyun Huang*

*State Key Laboratory of Powder Metallurgy, Central South University, Changsha, Hunan, P. R. China 410083.*

Email: hby@mail.csu.edu.cn. guoqin999@gmail.com

## Experimental Section:

The reaction equations for precursor precipitation are as follows:

$$Co^{2+} + xF^- \rightarrow CoF_x^{(x-2)-}$$

$$Ni^{2+} + xF^- \rightarrow NiF_x^{(x-2)-}$$

$$H_2NCONH_2 + H_2O \rightarrow 2NH_3 + CO_2$$

$$CO_2 + H_2O \rightarrow CO_3^{2-} + 2H^+$$

$$NH_3 \bullet H_2O \rightarrow NH_4^+ + OH^-$$

$$NiF_x^{(x-2)-} + 2CoF_x^{(x-2)-} + (3-1.5y)CO_3^{2-} + 3yOH^- + nH_2O$$
$$\rightarrow NiCo_2(OH)_{3y}(CO_3)_{(3-1.5y)} \bullet nH_2O + 3xF^-$$

After heat treatment, $NiCo_2O_4$ was obtained according to the following equation:

$$2NiCo_2(OH)_{3y}(CO_3)_{(3-1.5y)} \bullet nH_2O + O_2$$
$$\rightarrow 2NiCo_2O_4 + (2n+3y)H_2O + (6-3y)CO_2$$

## Results and Discussion Section:

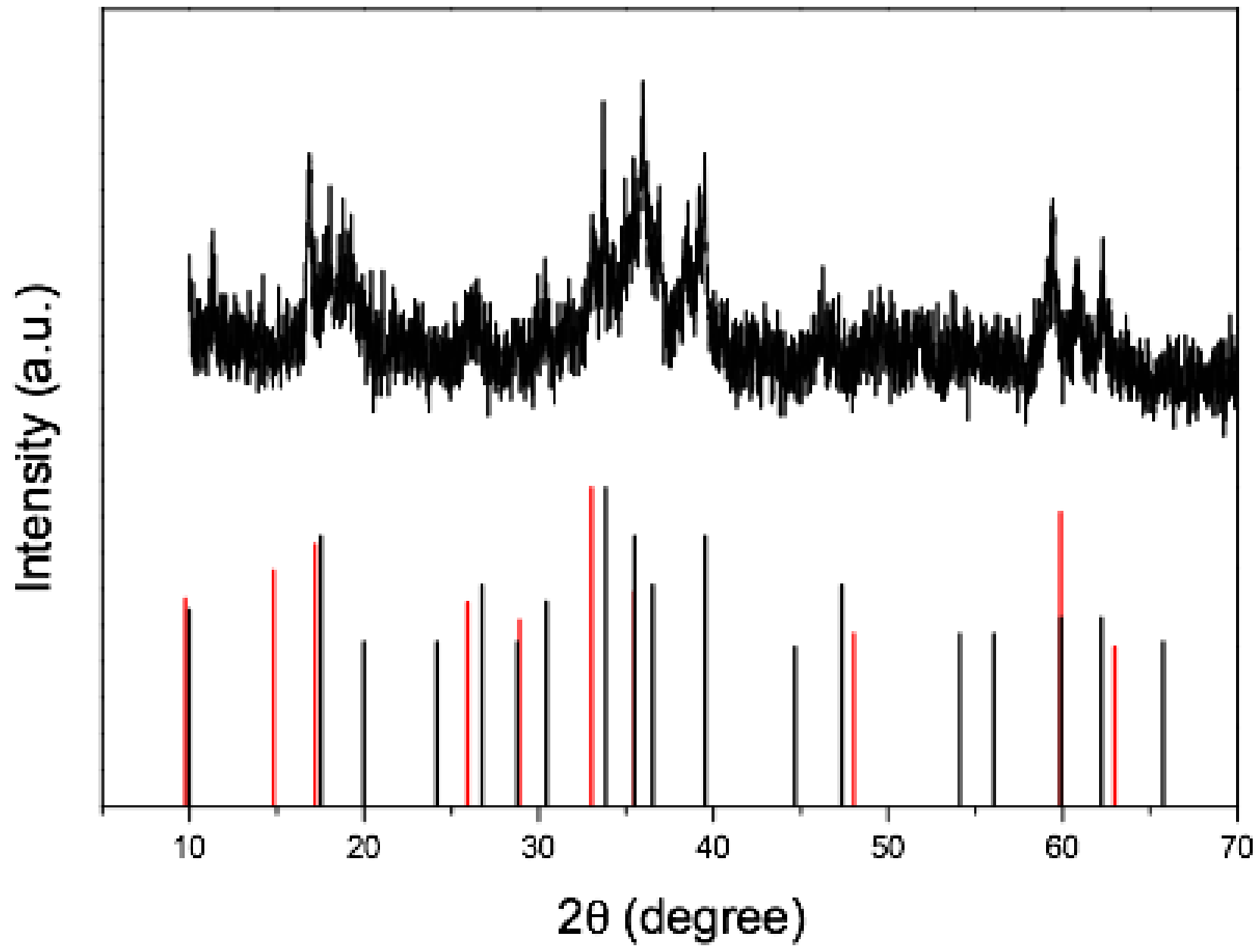

Fig. S 1 XRD patterns of precursor powders by hydrothermal method (red mark line for $Ni_2(OH)_2CO_3 \cdot 4H_2O$; black marline for $Co(CO_3)_{0.5}(OH) \cdot 0.11H_2O$ )

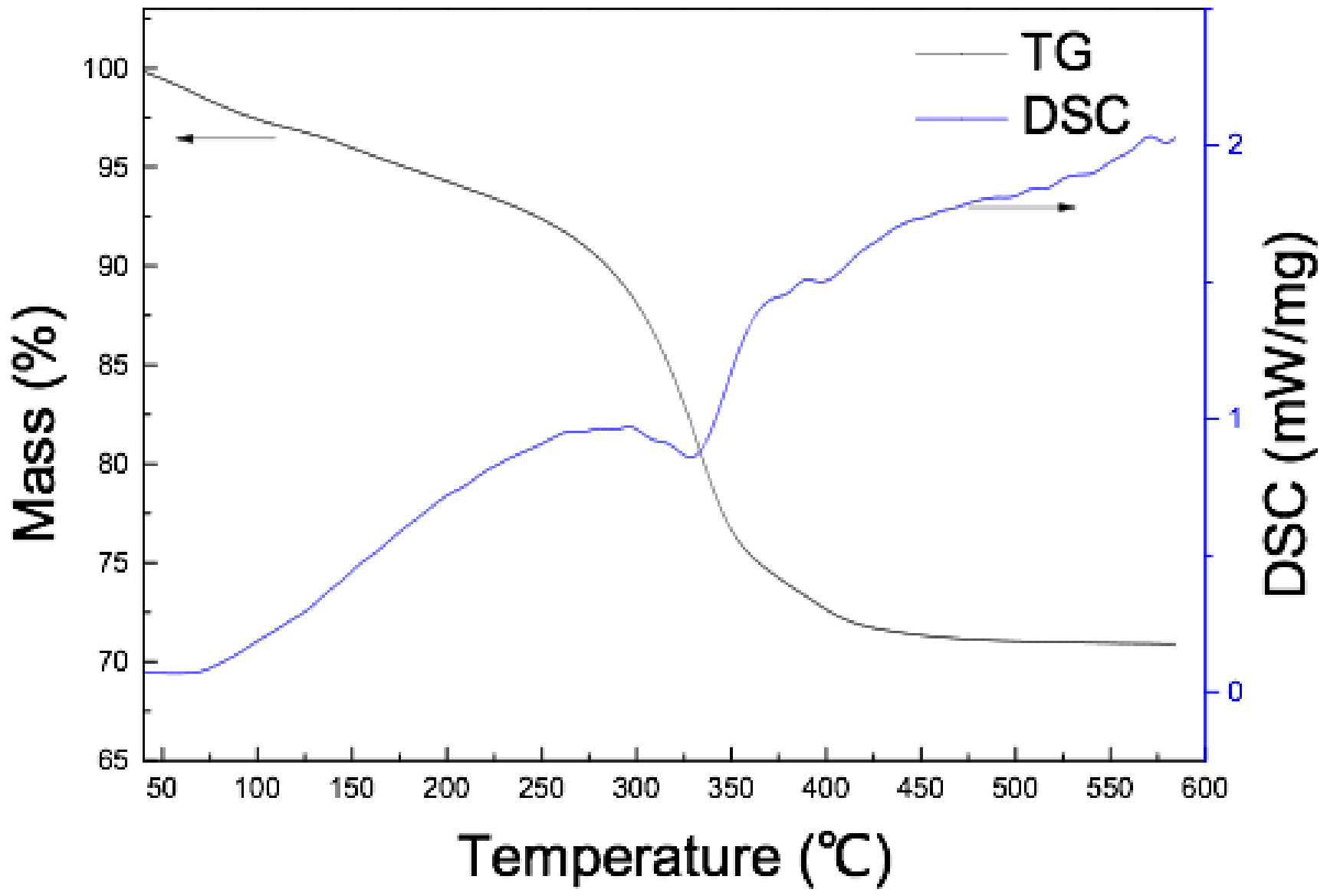

Fig. S 2 The thermal gravimetric analysis and differential scanning calorimetry curves of precursors by hydrothermal method

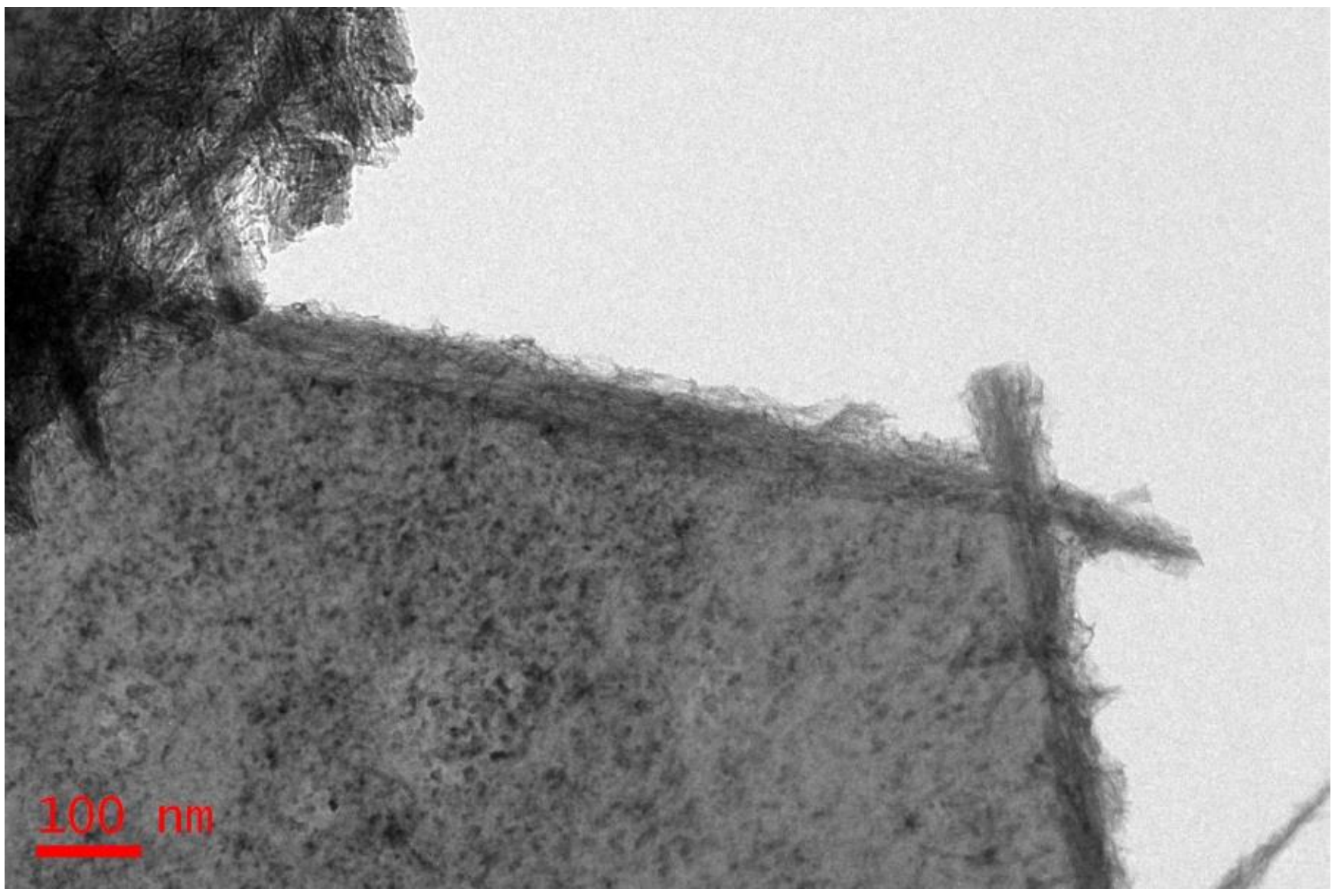

Fig. S 3 TEM micrographs of precursors by hydrothermal method before annealing

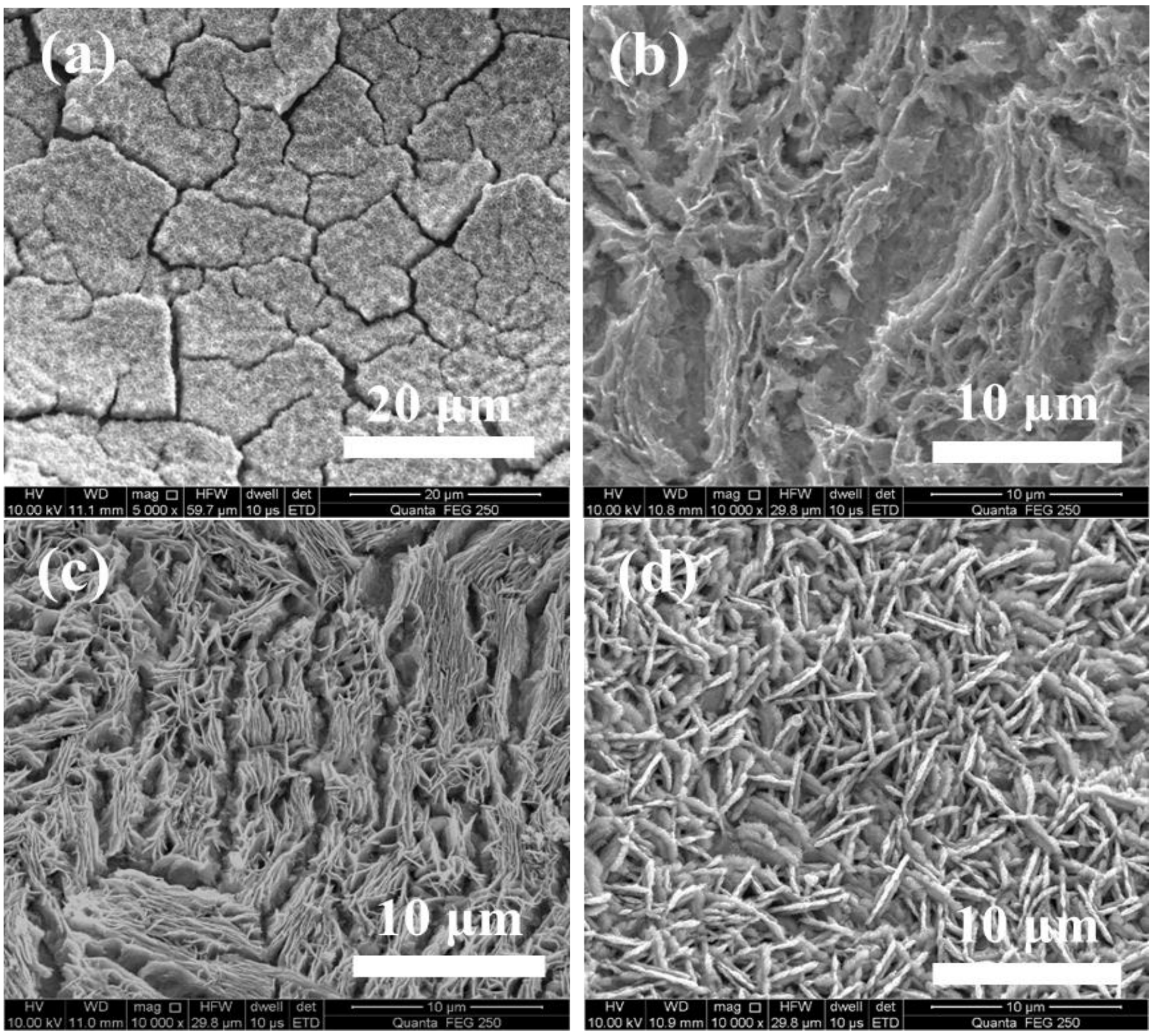

Fig. S 4 After cycling SEM micrographs of $NiCo_2O_4$ synthesized with varying combination of $NH_4F$ dosages and hydrothermal reaction time (a) (2 mmol, 10h) (b) (6 mmol, 5h) (c) (12 mmol, 3h) (d) (24 mmol, 1h)

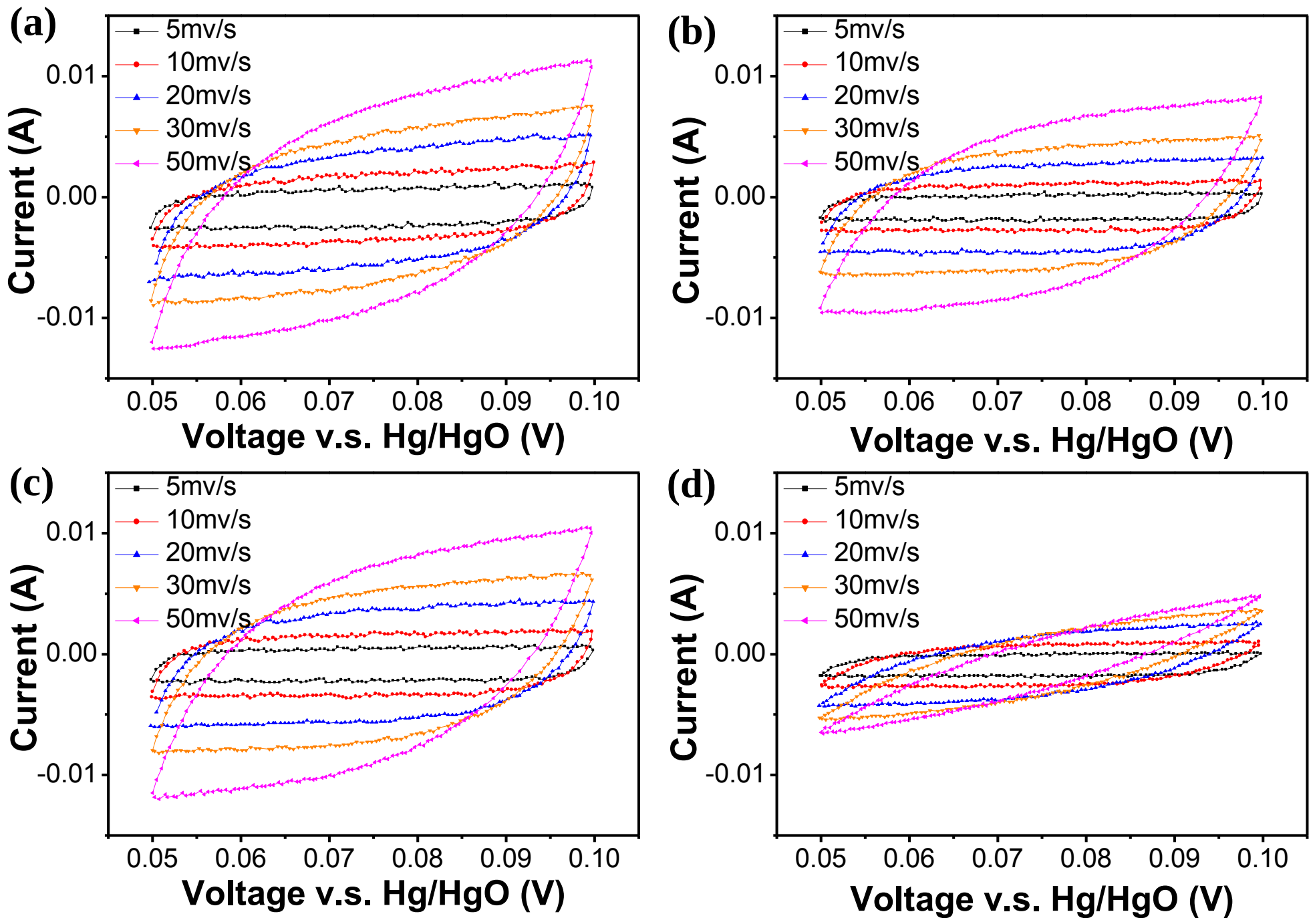

Fig. S 5 CV curves of $NiCo_2O_4$ with different micro-structures between 0.05 and 0.1V v.s. Hg/HgO at different scan rates (a) NWs (b) V-NPs@NWs (c) V-NPs (d) H-NPs

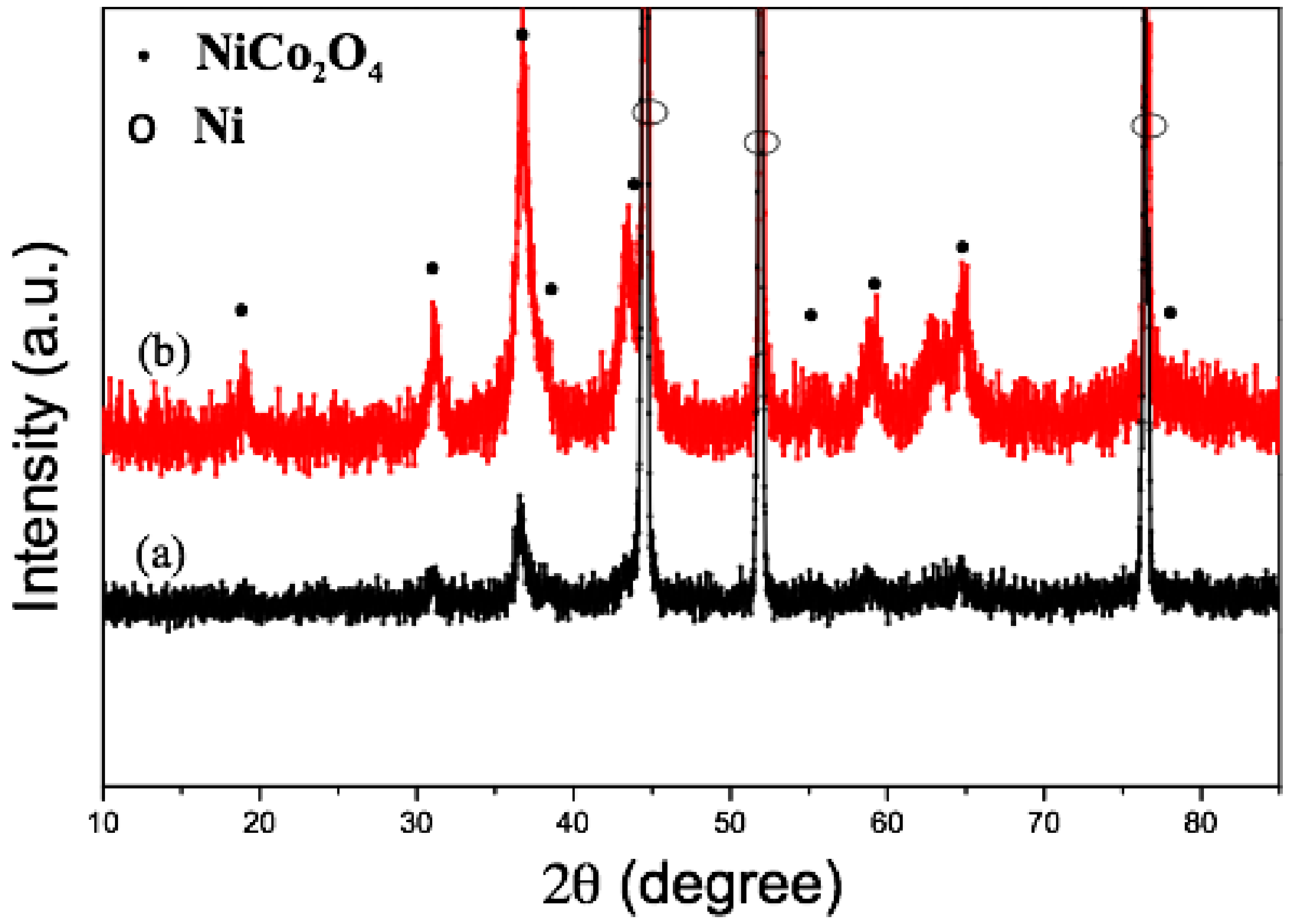

Fig. S 6 XRD patterns of Ni@$NiCo_2O_4$ electrodes on different conductive substrates (a) MP Ni (b) HP Ni.

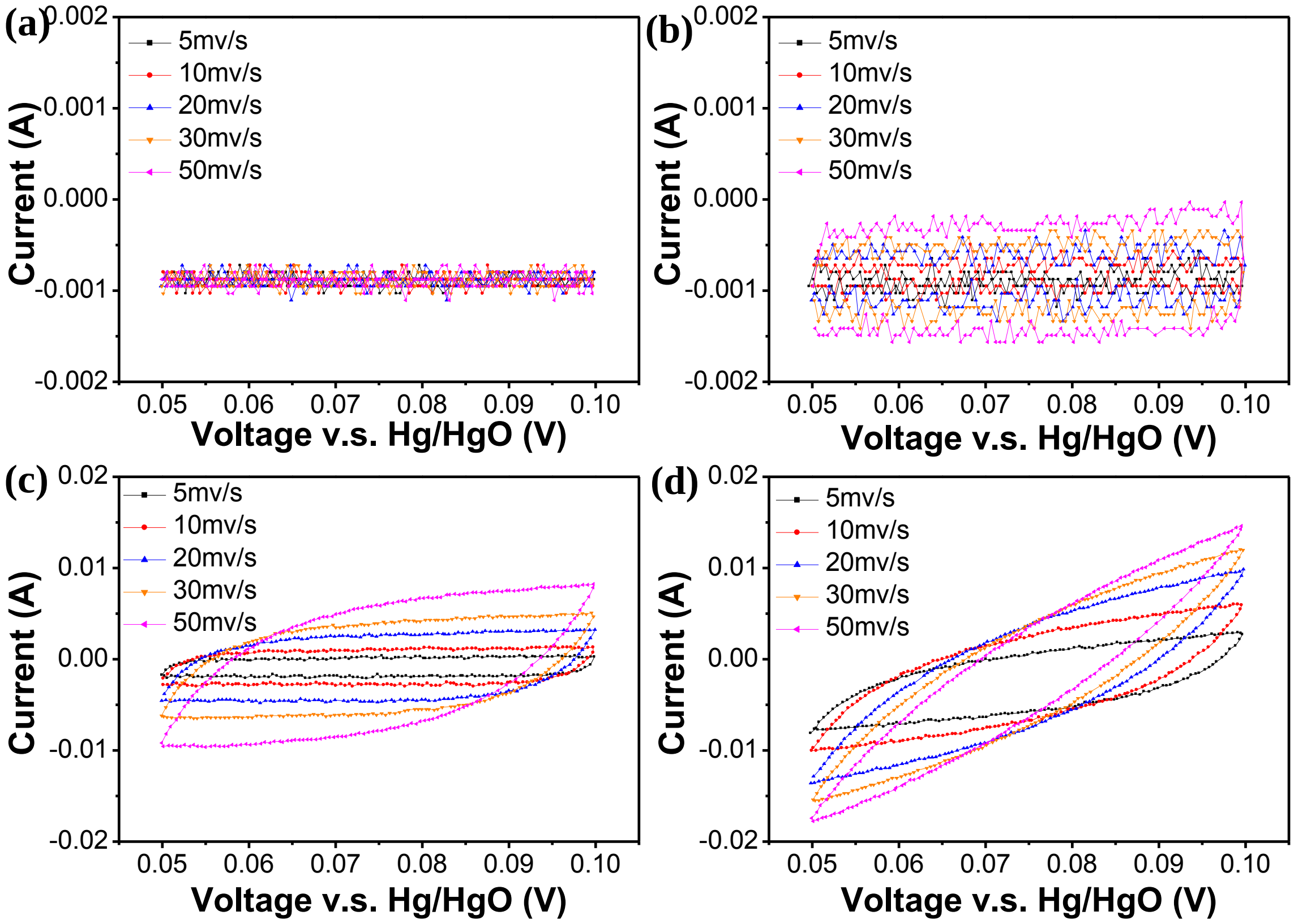

Fig. S 7 CV curves of porous Ni substrates and porous Ni@$NiCo_2O_4$ between 0.05 and 0.1V v.s. Hg/HgO at different scan rates (a) MP Ni (b) HP Ni (c) MP Ni@$NiCo_2O_4$ (d) HP Ni@$NiCo_2O_4$

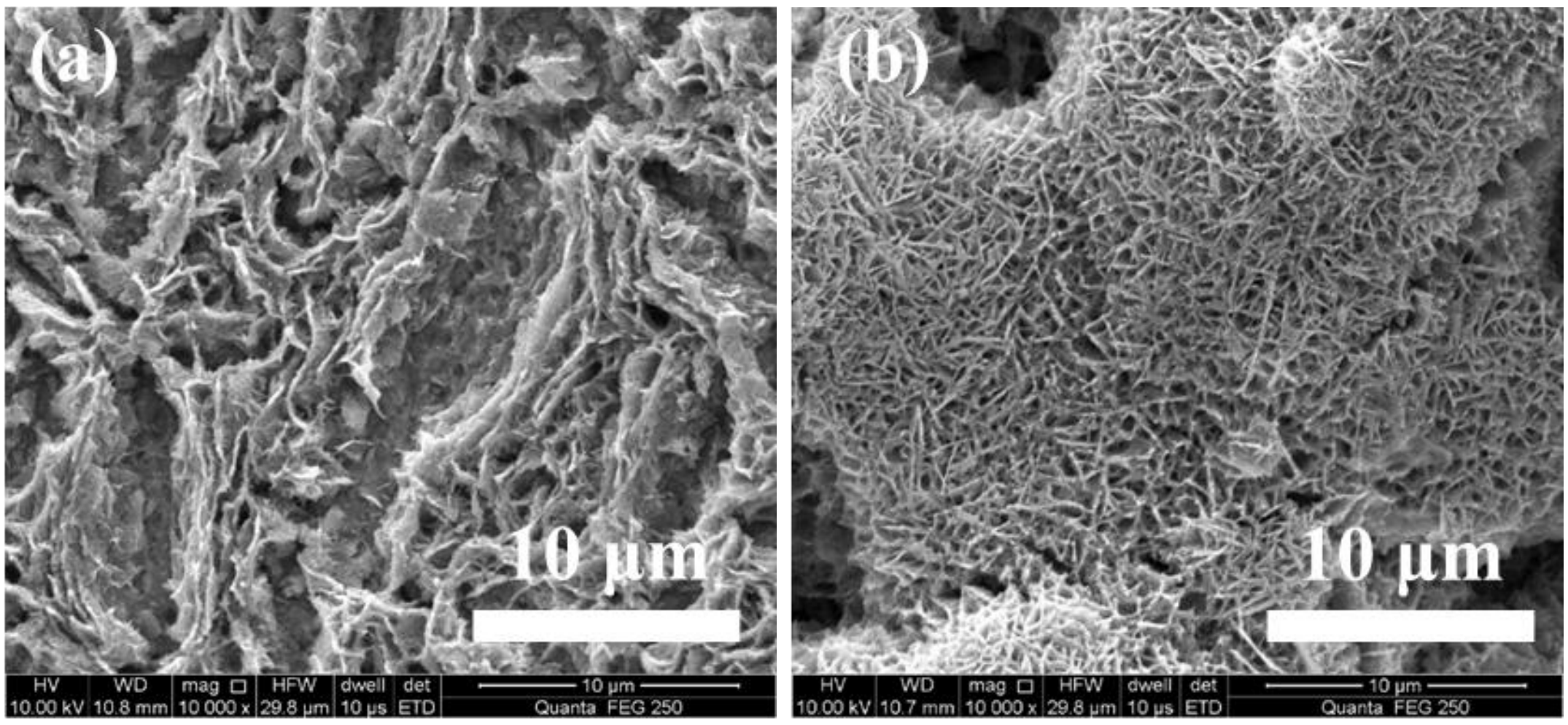

Fig. S 8 After cycling SEM micrographs of $NiCo_2O_4$ on different conductive substrates after cycling test (a) on MP Ni (b) on HP Ni